\documentclass{aa}
\usepackage{graphicx}
\newcommand{\msun}{{M$_\odot$}\ }
\def\aua{\rm A\&A}

\def\apj{\rm ApJ}
\def\aj{\rm AJ}

\def\mnras{\rm MNRAS}

\begin{document}

\title{(Sub)millimetre emission from \object{NGC~1569}: an abundance 
of very small grains}

\titlerunning{Submillimetre and millimetre emission of \object{NGC~1569}}
\author{U. Lisenfeld \inst{1}
\and F.P. Israel \inst{2} \and J.M. Stil \inst{2,}\inst{3}\and A. Sievers 
\inst{1}}

\institute{IRAM, Avendida Divina Pastora 7, N.C., 18012 Granada, Spain
\and Sterrewacht Leiden, Postbus 9513, 2300 RA Leiden, The Netherlands
\and Physics Department,  Queen's University, Kingston ON K7L 4P1, Canada}

\date{Received   / Accepted   }

\abstract{
We present new data of the dwarf galaxy \object{NGC~1569} at 450 $\mu$m, 850 $\mu$m 
and 1200$\mu$m taken with SCUBA at the JCMT and the bolometer array at the 
IRAM 30m telescope.
After including data from IRAS at 12, 25, 60 and 100 $\mu$m, we have 
successfully fitted the dust grain population model of D\'esert et al. 
(1990)
to the observed midinfrared-to-millimeter spectrum. The fit requires a 
combination of both large and very small grains exposed to a strong 
radiation field as well as an enhancement of the number 
of very small grains relative to the number of large grains. We interpret 
this as the consequence of large grain destruction due to shocks 
in the turbulent  interstellar medium of \object{NGC~1569}. The contribution
of polyaromatic hydrocarbons (PAH's) is found to be negligible.
Comparison of the dust emission maps with an HI map of similar resolution
shows that both dust and molecular gas distributions peak close to the
radio continuum maximum and at a minimum in the HI distribution.
From a comparison of these three maps and assuming that the gas-to-dust mass
ratio is the same everywhere, we estimate the
ratio of molecular hydrogen column density to integrated CO intensity
to be about 25 -- 30 times the local Galactic value. The gas-to-dust
ratio is 1500 -- 2900, about an order of magnitude higher than in the Solar
Neighbourhood. 
}

\maketitle

\section{Introduction}

Dwarf galaxies characteristically have low metallicities and consequently 
low dust abundances. In dwarf galaxies, both  dust properties and
amounts differ from those in spiral galaxies, as for instance implied by 
a difference in IRAS colours (cf. Melisse \& Israel \cite{melisse}). 
However, actual 
dust abundances and dust composition are only poorly known for dwarf 
galaxies. The dust mass of a galaxy can be estimated reliably only if 
good (sub)mm measurements allow to constrain the amounts of relatively 
cold dust which may dominate the total dust mass with only a very 
limited contributions to the emission at infrared wavelengths. Such data 
are available for a limited number of galaxies, and are especially scarce 
for faint dwarf galaxies.

\object{NGC~1569} (Arp~210; VII~Zw~16) is a nearby irregular dwarf galaxy at a 
distance of 2.2 Mpc (Israel 1988,  
hereafter \cite{israel88}). It is a member of the 
low-galactic latitude IC~342/Maffei~1/Maffei~2/Dw~1 group, containing at 
least 15 dwarf galaxies (Huchtmeier et al. \cite{huchtmeier}). 
\object{NGC~1569} is presently 
in the aftermath of a massive burst of star formation 
(\cite{israel88}; Israel \& De Bruyn \cite{israel-bruyn}; Waller 
\cite{waller}).
{ 
Its present star formation rate, derived from the H$\alpha$ luminosity,
is 0.4 \msun yr$^{-1}$ (Waller \cite{waller}).  In the recent past,
this galaxy has experienced a starburst 
which started about $1 - 2 \times 10^7$ yr ago as estimated by Israel
(\cite{israel88}) and about $1.5 \times 10^7$ yr ago determined by
Vallenari \& Bomans (\cite{vallenari}) from colour-magnitude diagrams.
The end of the starburst, about 5 Myr ago, can be well dated by a kink in the
synchrotron spectrum  (Israel \& De Bruyn \cite{israel-bruyn}) and 
photometric studies
(Vallenari \& Bomans \cite{vallenari}, Greggio et al \cite{greggio}).
%
%
}
Although a small galaxy with neutral atomic 
hydrogen (HI) dimensions of 3 x 2 kpc
(Israel  \& Van Driel \cite{israel90}; Stil \& Israel,  
in preparation, hereafter \cite{stil01}), \object{NGC~1569} contains 
two extremely compact luminous star clusters A and B 
(Ables \cite{ables}; Arp \& Sandage \cite{arp}; Aloisi et al. \cite{aloisi} 
and references therein) with 
bolometric luminosities of order 10$^{8}$ L$_{\odot}$ located in a 
deep HI minimum. A third such cluster is embedded in bright emission 
nebulosity 
(F.P. Israel $\&$ W. Wamsteker, unpublished; Prada et  al. \cite{prada}; ;
listed as No. 10 by Hunter et al. \cite{hunter}) coincident with the peak of 
the radio continuum distribution. 

The HI observations show the galaxy to be in solid-body rotation out to a 
radius of 1$'$ (0.64 kpc); however, much of the HI is in chaotic motion 
(\cite{stil01}). 
To the east of \object{NGC~1569}, an apparently counterrotating HI cloud 
connected by an HI bridge to \object{NGC~1569} is observed (Stil $\&$ Israel 
\cite{stil98}). 
The brightest HI with column densities N(HI) $\approx 2 \times 10^{20}$ 
cm$^{-2}$, occurs in the three main peaks of an HI ridge in the northern 
half of the galaxy. About a third  of the total mass of \object{NGC~1569} 
resides 
in its neutral atomic hydrogen (\cite{israel88}, \cite{stil01}). 
Weak CO emission is found
close to the radio continuum, between two HI maxima. It was mapped in the 
$J$=1--0 and $J$=2--1 transitions by Greve et al. (\cite{greve}). 
Comparison with 
the $J$=1--0 CO detection in a larger beam by Taylor et al. 
(\cite{taylor98}) suggests 
that the maps by Greve et al. contain virtually all CO emission from the 
western half of \object{NGC~1569}. Aperture synthesis maps of the same two CO 
transitions with the IRAM interferometer show distinct CO clouds of sizes 
ranging from 40 to 100 pc (Taylor et al. \cite{taylor99}). 
As these maps recover only 
a quarter of the single-dish flux, weak and more diffusely distributed CO 
must be present. CO(3-2) observations  
show a high $J$=3--2/$J$=2--1 ratio of 1.4
(M\"uhle et al. \cite{muehle}) indicating a warm molecular gas phase.

The far-infrared emission from \object{NGC~1569}, observed with IRAS, is 
remarkably
strong for a dwarf galaxy; the continuum spectrum indicates that the dust
in the galaxy is exposed to intense radiation fields (\cite{israel88}).

In this paper we extend the far-infrared spectrum of \object{NGC~1569} to 
(sub)millimeter wavelengths by presenting new observations obtained with 
the IRAM and JCMT telescopes. These new observations allow us, for the
first time, to determine the amount of dust in this galaxy and to study 
its properties in detail.

\section{Observations}

Observations at 850$\mu$m and at 450 $\mu$m were made in October 1997 and 
again in March 2000 with  the SCUBA camera at the JCMT on Mauna Kea 
(Hawaii). As the quality of the later observations greatly surpasses
that of the earlier, we only discuss the March 2000 observations in 
this paper. SCUBA consists of two bolometer arrays of 91 elements at 450 
$\mu$m and 37 elements at 850 $\mu$m, both with a field of view of about 
2.3$'$ (Holland et al. \cite{holland}), thus covering the whole galaxy. Both 
wavelengths were observed simultaneously in jiggle mode. The total 
integration  times, half on source, half on sky, was 5 hrs. 
The images were chopped with a throw of 120$''$ at a frequency of 7.8 Hz.  

We applied the standard reduction procedure: this includes flat-fielding, 
removal of transient spikes, correction of atmospheric opacity 
($\tau_{850} \approx 0.3; \tau_{450} \approx 1.1$); pointing correction and 
sky-removal. The data were calibrated by observation of the standard 
source CRL 618 ($F_{850} = 11.2$ Jy, $F_{450} = 4.56$ Jy) immediately 
before and after the \object{NGC~1569} observations. From the 
calibrator images, 
we determined a beam-size (FWHM) of 15.3 $\times$ 15.6 $''$ at 850 $\mu$m.
 
Observations at 1200$\mu$m were made in December 1998 at the IRAM 30m 
telescope on Pico Veleta (Spain), with the 19-channel bolometer array
of the Max-Planck-Institut f\"ur Radioastronomie (MPIfR). Additional 
maps at 1200$\mu$m were obtained in April 1999 and again in March 2000 
with the 37-channel bolometer of the MPIfR. The beam-size is 10.8\arcsec.
The observations were done on-the-fly, with a wobbler throw of 46\arcsec. 
Opacities ranged from 0.1 to 0.3. The data were reduced in a standard 
manner 
including baseline subtraction, spike 
removal and sky noise removal. The maps were calibrated by observation 
of the planets Mars and Uranus.

\section{Integrated flux-densities} 

Maps of \object{NGC~1569} at the three observed wavelengths are
 shown in Fig.~1 and 2. 
We have optimized the SCUBA dataset by excluding the two samples out of 
ten that were taken under poor atmospheric opacity conditions.
{ The noise levels of the maps are 2.5 mJy/beam (1200 $\mu$m), 
4 mJy/beam (850 $\mu$m) and 68 mJy/beam (450 $\mu$m).}

We determined total galaxy flux-densities by integrating the maps 
over increasingly larger areas until the cumulative flux-densities thus
obtained converged to a final value. However, it was found that low-level
emission from \object{NGC~1569} extended over most of the SCUBA field of view
rendering an accurate determination of the (sky) zerolevel questionable.
This was not a problem in the larger IRAM 1200$\mu$m field. 
For this reason, we fitted the 850$\mu$m and 450$\mu$m
cumulative growthcurves to the scaled growthcurve at 1200$\mu$m, 
allowing for a zerolevel offset of the 850$\mu$m and 450$\mu$m
growthcurves. The scaling of the 1200$\mu$m growthcurve was uniquely
defined by the requirement that the difference between the 
growthcurves increases as the square of the radius, as expected if
the SCUBA maps had a zerolevel offset.

In this way, we obtained total flux-densities S$_{1200}$ = 250$\pm$30 mJy,
$S_{850}$ = 410$\pm$45 mJy and $S_{450}$ = 1820$\pm$700 mJy (Table 1). 
The errors quoted are formal errors based on map noise and including,
at 450$\mu$m and 850$\mu$m, the error introduced by the zerolevel correction.
The latter error is, however, small: From the goodness of the fit
we estimate it to be about 4\% at both 450$\mu$m and 850$\mu$m.
growthcurves. The actual 
uncertainty is larger because of calibration uncertainties. In particular 
at 450 $\mu$m, the telescope error pattern and relatively high atmospheric 
opacities conspire to produce a large uncertainty at this wavelength. We 
estimate the total error, including
opacity correction, calibration, noise of the map and zerolevel
correction to be about 
50\% at 450 $\mu$m and 30\% at both 850 $\mu$m and 1200$\mu$m.
\begin{figure}
\resizebox{\hsize}{!}{\rotatebox{270}{\includegraphics{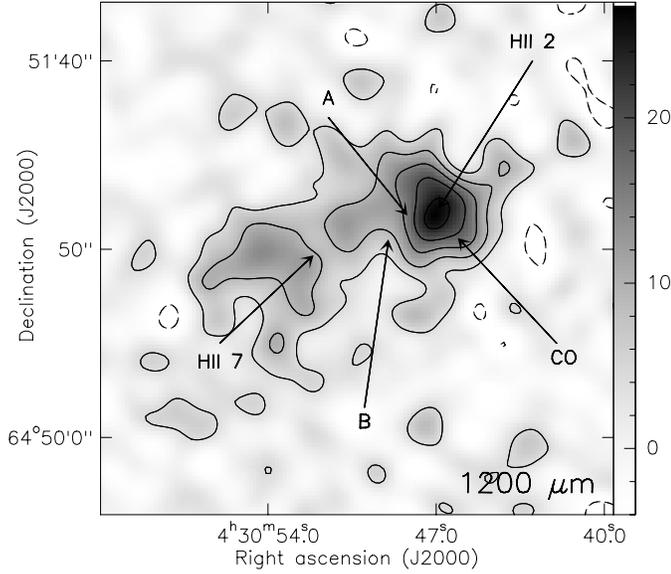}}}
\caption{Map of \object{NGC~1569} at 1200 $\mu$m, 
smoothed to a resolution of 13\arcsec. Contour values
are at levels -n,  n, 2n, 3n ...., where n = 5 mJy/beam.
The lowest contour corresponds to 2$\sigma$, { where $\sigma$
is the noise level of the map.
The positions of the starclusters A and B
are indicated, as well as the maximum of the CO
distribution (Tayler et al. \cite{taylor98}) and the position of the
two most prominent HII regions, catalogued as number 2 and 7 by
Waller (\cite{waller})}.
}
\end{figure}
\begin{figure}
\resizebox{\hsize}{!}{\includegraphics{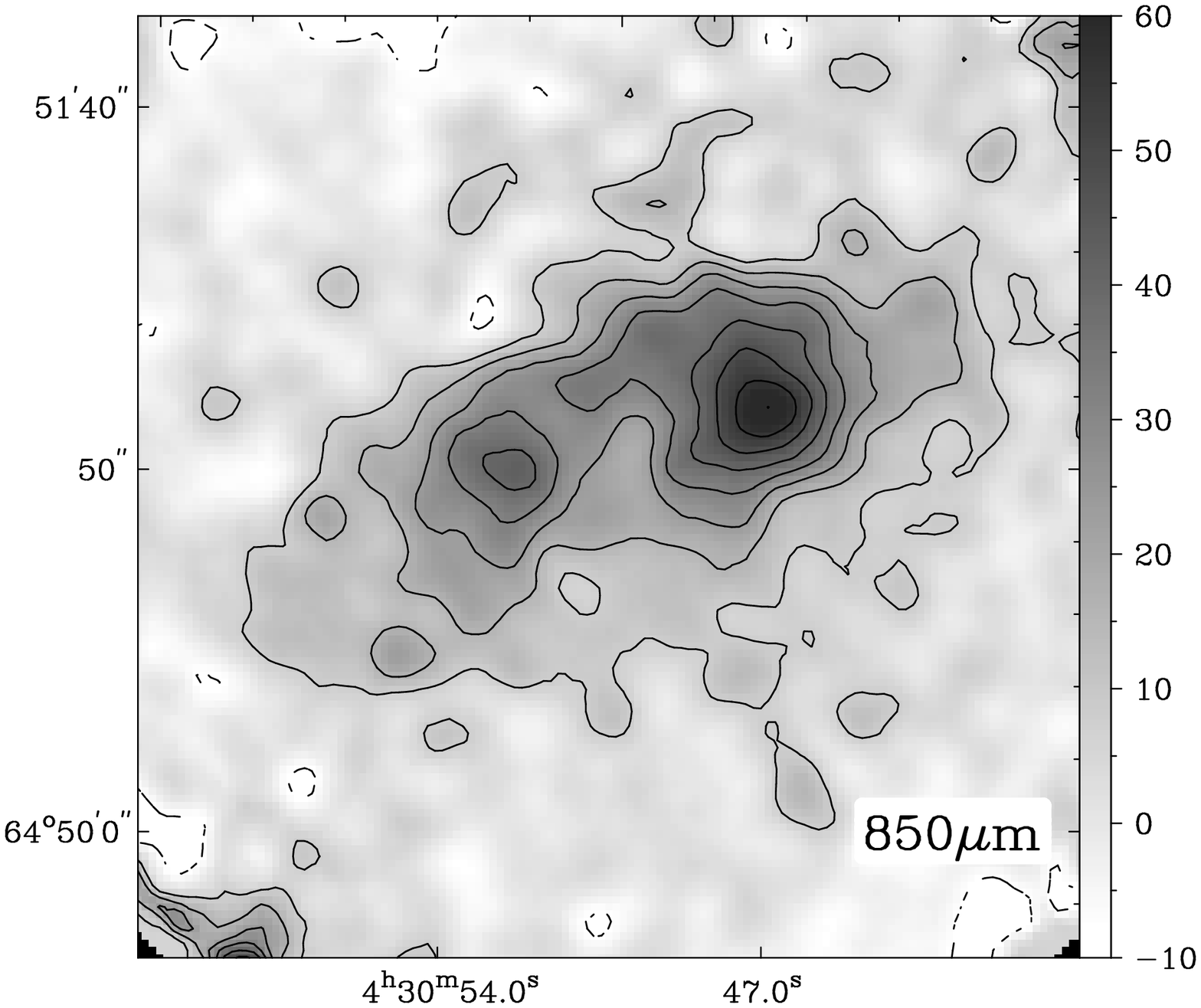}}
\resizebox{\hsize}{!}{\includegraphics{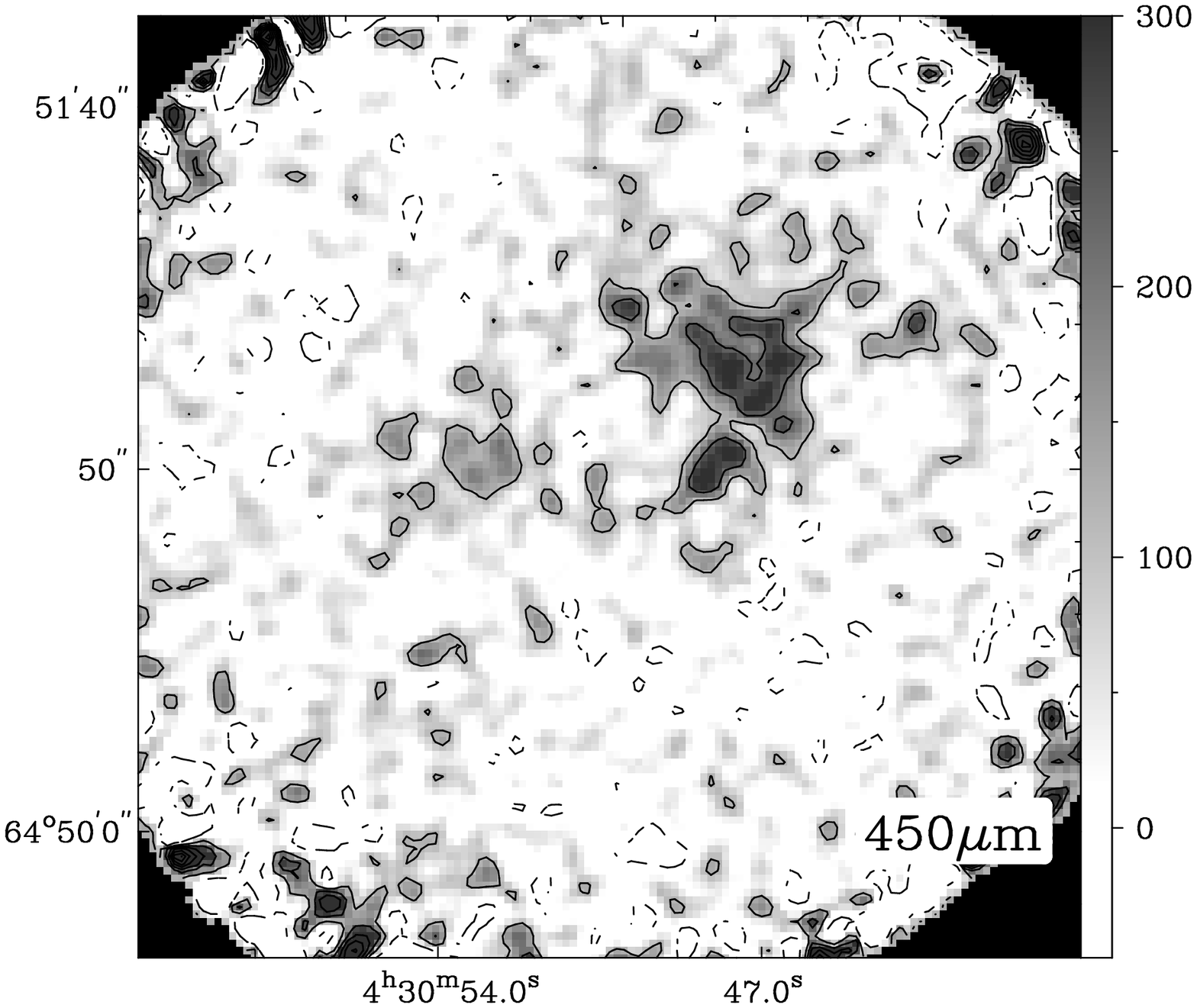}}
\caption{Maps of \object{NGC~1569} at 850
and 450 $\mu$m. Contour values
are at levels -n,  n, 2n, 3n ...., where 
n = 8 mJy/beam (850 $\mu$m) and n = 115 mJy/beam (450$\mu$m).
The lowest contours correspond to 2$\sigma$, { where $\sigma$
is the noise level of the map.}
%
}
\end{figure}

Especially at the longer wavelengths of 850 $\mu$m and 1200 $\mu$m, the 
broad-band flux-densities determined from the maps may contain 
non-negligible contributions by thermal free-free continuum emission 
and CO line emission. The contribution of the CO line to 1200$\mu$m 
broadband emission can be estimated 
from the $J$=2--1 CO measurements by Greve et al. (\cite{greve}). 
Adding up all the 
emission detected by them, we obtain $I_{\rm CO}=13.6$ K km s$^{-1}$ at 
230 GHz. 
Allowing for the presence of a similar amount of undetected CO emission in
the remainder of \object{NGC~1569} and using a conversion factor between
$T_{\rm mb}$ and flux density of 4.6 Jy/K and an adopted instrumental 
bandwidth of 50 GHz for the bolometer, we obtain a maximum of 2.1 mJy 
contributed by the $J$=2--1 CO line to the total emission in the 1200 
$\mu$m band. The $J$=3--2 CO emission, contributing to the 850$\mu$m 
broadband emission, has been measured by Meier et al. (\cite{meier}). 
At the CO 
peak in the map by Greve et al. (\cite{greve}), 
they find the $J$=3--2/$J$=1--0 
ratio to be unity, characteristic of rather hot and dense molecular gas. 
From this result, we estimate a maximum contribution of 10 mJy to the 
emission in the 70 GHz wide SCUBA band at 850 $\mu$m. We thus 
find that, at either wavelength, contributions by CO to the total emission 
of \object{NGC~1569} are unimportant, reflecting the general weakness 
of CO emission 
from dwarf galaxies.

Consideration of the thermal free-free continuum contribution does not
warrant a similar conclusion. Its intensity has been estimated from radio 
maps at 1.5, 5, 8.4 and 15 GHz by Wilding (\cite{wilding}). The derived total 
{\it thermal} radio continuum flux-density at 1 GHz is 100 mJy, in very
good agreement with the value of 97 mJy estimated by Israel \& De Bruyn 
(\cite{israel-bruyn}) from reddening-corrected H$\alpha$ measurements. 
Scaling this 
emission to higher frequencies (shorter wavelengths) as $\nu^{-0.1}$, we 
derive considerable thermal free-free contributions to the total 
emission of 58, 56 and 52 mJy at 1200, 850 and 450 $\mu$m (23$\%$, 13$\%$ 
and 3$\%$) respectively. In Table~1, we have summarized these results,
{ together with further photometric data from the literature.}
\begin{table}
\caption{\object{NGC~1569} emission from UV to millimetre}
\begin{flushleft}
\begin{tabular}{rcl}
\hline
\noalign{\smallskip}
    	&  \multicolumn{2}{l}{UV and optical flux density} \\ 
$\lambda$ & flux density & Ref. \\
(\AA)  &$10^{-14}$ erg s$^{-1}$ cm$^{-1}$ \AA$^{-1}$ & \\
\hline
1500 & 152 $\pm$ 48 & (1) \\
1800 & 208 $\pm$ 25 & (1)\\
2200 & 146 $\pm$ 25 & (1)\\
2500 & 100 $\pm$ 18 & (1)\\
3650 & 108 $\pm$ 10 & (2)\\
4400 & 101 $\pm$ 9& (2)\\
5500 & 70 $\pm$ 6& (2)\\ 
\end{tabular}

\begin{tabular}{rrrrr}
\hline
\noalign{\smallskip}
     	&  \multicolumn{4}{c}{Integrated flux density  
from 12$\mu$m to 1200$\mu$m} \\
$\lambda$ & Observed & Thermal+CO & Dust & Ref.\\
$(\mu$m) & (Jy) & (Jy)  & (Jy) & \\ 
\hline
12   	&  0.59	& ---   &  0.59$\pm$ 0.06  & (3)\\
25   	&  5.95 & ---   &  5.95 $\pm$ 0.6 & (3)\\
60   	& 44.6  & ---   & 44.6 $\pm$ 4.5 & (3)\\
100  	& 52.2  & ---   & 52.2 $\pm$ 5.2 & (3)\\
155     & 36    & ---   & 36 $\pm$ 11 & (4) \\
450  	&  1.82  & 0.052 &  1.77 $\pm$ 0.88 & (5)  \\
850  	&  0.41 & 0.066 &  0.34 $\pm$ 0.10& (5)\\ 
1200 	&  0.25 & 0.060 &  0.19 $\pm$ 0.06 & (5)\\ 
\hline
\end{tabular}
\end{flushleft}
(1) \cite{israel88}, corrected for Galactic foreground extinction
as described therein.

(2) De Vaucouleurs et al. (\cite{RC3}), corrected  
for Galactic foreground extinction based on $E(B-V)$ (\cite{israel88})
and a standard solar neighbourhood reddening law as in  \cite{israel88}.

(3) IRAS point 
source catalogue, flux densities are colour-corrected as described therein.

(4) Hunter et al. (\cite{hunter89})

(5) This work
\end{table}

\section{Nature of the dust emission}

\subsection{Observed emission spectrum}

The dust emission and extinction in our Galaxy can be modelled as the
combined effect of three components (e.g. D\'esert et al. 
1990 -- hereafter
\cite{desert}; 
Siebenmorgen \& Kr\"ugel \cite{siebenmorgen}). 
These components are: ({\it i}) large 
grains obeying a power-law size distribution (cf. Mathis et al. 
\cite{mathis77}), 
({\it ii}) polyaromatic hydrocarbons (PAH's) responsible for mid-infrared 
continuum and spectral feature emission, ({\it iii}) very small grains 
(VSG's) with sizes of a few nm.
The large grains are in equilibrium with the radiation field and their 
emission $F_\lambda$ can be described by a modified black-body curve 
following from the Kirchhoff-law: $F_\lambda = B(\lambda,T) k_\lambda$, 
where $B(\lambda,T)$ is the Planck function, $\lambda$ the wavelength, 
$T$ the dust temperature  and $k_\lambda$ the extinction coefficient.
In contrast, the VSG's are not at all in equilibrium. Their
heat capacity is so limited that they can be heated to very high 
temperatures ($T\approx 1000$ K) by the absorption of a single photon.
However, as they cool down rapidly, an ensemble of small grains cannot 
be described by a single temperature but rather by a relatively 
broad temperature distribution. Consequently, its integrated 
emission spectrum is significantly broader than that of large grains
with a limited temperature range. Because the small-grain temperature is 
on average higher, this spectrum peaks in the mid-infrared.
The wavelength dependence of $k_\lambda$ is not well known, however, models of 
interstellar dust that are both astronomically realistic in composition and
able to reproduce the observed dust extinction and emission curves,
predict $\beta \simeq 2$ for large grains (e.g. Draine \& Lee \cite{draine},
Ossenkopf \& Henning \cite{ossenkopf}, 
Kr\"ugel \& Siebenmorgen \cite{kruegel94}).
Observations of actively star-forming galaxies, for which the
dust emission beyond about 60 $\mu$m can be well described by
only one temperature component, confirm a value of $\beta$ close to 2
(e.g. Chini et al. \cite{chini}, Kr\"ugel et al \cite{kruegel98}).
{ A more general result has been obtained by Dunne et al. (\cite{dunne}) 
who have derived on a statistical basis 
$\beta \simeq 2 $ 
for a sample of nearby galaxies with 450 and 850 $\mu$m data.} 
For small amorphous grains, a lower value of $\beta \simeq 1$
can perhaps be expected (Seki \& Yamamoto \cite{seki}, 
Tielens \& Allamandola \cite{tielens}).

The integrated spectrum from 12$\mu$m to 1200$\mu$m of \object{NGC~1569}
provides important clues for the origin of 
the emission and for the relative importance of each of the three components
described above. Flux-densities at wavelengths $\lambda < 100\mu$m define 
a slope much less steep than expected for a modified black-body curve. 
In particular flux-densities at 12$\mu$m and 25$\mu$m are much higher than 
expected for such a curve. Moreover, the steep rise in intensity from 12 
to 25 $\mu$m precludes any significant contribution by emission from PAH's, 
as the broadband spectrum of these {\it decreases} with wavelength. At the 
long-wavelength side, the corrected flux-densities between 450 and 1200 
$\mu$m are nominally proportional to about $\lambda^{-2.5}$, i.e. proportional 
to a modified blackbody emission with a wavelength dependence $\beta \leq 1$, 
more characteristic for  small than for large dust grains, or
typical for a superposition of various large grain temperature components.

In order to further analyze the dust population of \object{NGC~1569}, 
represented by 
the observed emission, we have modelled the spectrum for a few representative
cases. We present here two particular cases: (i) emission dominated by
large particles at different temperatures and (ii) a three-component fit 
in which the relative fractions of different types of dust particles are
varied. 

\begin{table}
\caption{Flux densities of the various dust components}
\begin{flushleft}
\begin{tabular}{llll|ll}
\hline
\noalign{\smallskip}
     	  \multicolumn{4}{c}{Model A: Cold dust} &
\multicolumn{2}{c}{Model B: Very small grains} \\
$\lambda$ & $S_{\rm hot}$ & $S_{\rm warm}$ & $S_{\rm cold}$& 
 $S_{\rm VSG}$ & $S_{\rm large\ grains}$ \\
$(\mu$m) & (Jy) & (Jy)  & (Jy) & (Jy)  & (Jy) \\ 
\hline
12 &  0.68  & ---   &  ---  & 0.44 & --   \\ 
25 &  5.7   &  0.18 &  ---  & 6.07  & 0.16 \\ 
60 &  1.9   &  41.2 &  ---  & 26.2 & 18.1 \\ 
100 & 0.46  & 52.5  &  ---  & 25.3 & 26.6 \\ 
155 & 0.11  & 26.6  &  0.02 & 17.2 & 14.1 \\ 
450 &  ---  & 1.1   &  0.70 & 1.9 & 0.54 \\ 
850 &  ---  &  0.12 &  0.26 & 0.39 & 0.05 \\ 
1200 & ---  & 0.03  &  0.11 & 0.16 & 0.01 \\ 
\hline
\end{tabular}
\end{flushleft}
\end{table}

\subsection{An abundance of cold dust?}

\begin{figure}
\resizebox{\hsize}{!}{\rotatebox{270}{\includegraphics{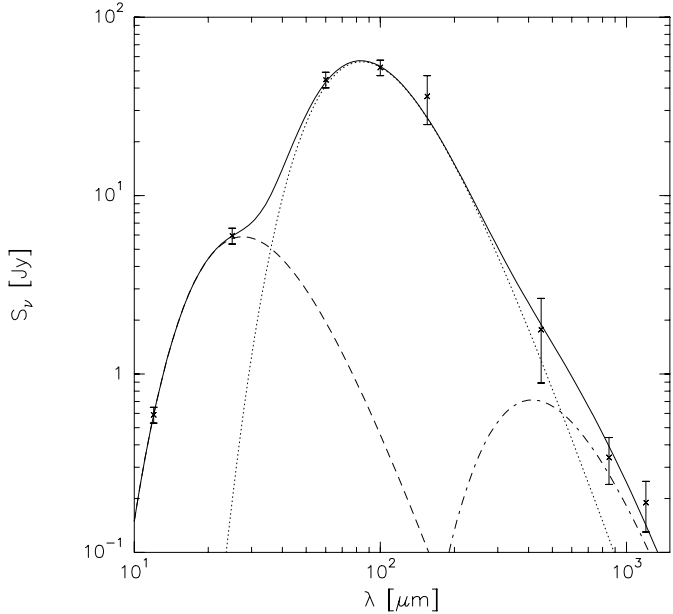}}}
\caption{The mid-infrared to millimeter spectrum of \object{NGC~1569}, fitted 
with
a three-temperature large-grain model. Crosses mark flux-densities from
Table 1. The three dust components have temperatures of 105 K (dashed line), 
34.5 K (dotted line) and 7 K (dashed dotted line).}
\label{threetemp}
\end{figure}

In principle, it is well possible to fit the emission spectrum over the
entire observed range with large grains at various temperatures. As we have
only seven observed points, we must limit ourselves to a model allowing for
three different temperature components. Obviously, even better fits 
can be obtained by allowing for more temperature components, but there
is no physical relevance to such an improvement. In Fig.~\ref{threetemp} 
we show the result, { adopting $\beta = 2$. Table 2 presents the flux
densities of the different components.}
The data can be fit very satisfactorily, which is no 
surprise given the large number of free parameters (almost equal to the
number of independent data points). As Fig.~\ref{threetemp} shows, each 
component is essentially determined by two different measured points. However,
in order to reproduce the specific features at the short and the long 
wavelength sides referred to in the previous section, the inclusion
of {\it both rather hot and rather cold dust} components is unavoidable.
The temperature of the coolest component required to fit the relatively 
high (sub)millimeter emission is very low, of the order of only 7 K.

Moreover, because of the low dust emissivities at this temperature, a very 
large fraction of {\it all dust} would have to be so cold in order to 
explain the emission at 850 $\mu$m and 1200$\mu$m. From the large grain 
component fits to the \object{NGC~1569} spectrum we may estimate the dust mass 
from each components 100$\mu$m emission:

\begin{equation}
M_d= \frac{D^2 S_{100 \mu m}}{B_{100 \mu m}(T) k_{100 \mu m}}
\label{mdust}
\end{equation}

\noindent
Here, $B_{100 \mu m}(T)$ is the Planck function at 100$\mu$m. 
We have assumed a dust extinction coefficient (per dust mass) of 
$k_{100 \mu m} = 63$ cm$^2$ g$^{-1}$. 
Extrapolating this value with $k_\lambda \propto \lambda^{-2}$ 
and adopting a gas-to-dust mass ratio of 150 we obtain  
$k_{1200 \mu m}=0.0025$ cm$^2$ g$^{-1}$ (per gas mass)
in good agreement with other studies
(Kr\"ugel et al. \cite{kruegel90}, Mezger et al. \cite{mezger90}, 
Draine $\&$ Lee \cite{draine}).
The results for the three components shown in Fig.~\ref{threetemp}
are a small mass
of hot dust, $M_{\rm d,h} = 12.4$ \msun, a significant mass of warm dust,
$M_{\rm d,w} = 3.1 \times 10^4$ \msun and a very large mass for the very cold
dust: $M_{\rm d,c} = 1.14 \times 10^6$ \msun. 
The total dust mass $M_{\rm d,tot}
=  1.17 \times 10^6$ \msun would thus be 
completely (i.e. for 97$\%$) dominated by 
the cold dust component.

{ How large is the uncertainty in this mass? 
The dust mass depends on the adopted dust temperatures and
wavelength dependence of the extinction coefficient, $\beta$.
The highest temperature of the cold dust that is just marginally
consistent with the errors of the 450 -- 1200 $\mu$m data  is
11 K. This temperature 
is mainly determined by the slope of the spectrum between 450 -- 1200
$\mu$m. With this, we derive a  total
dust mass of $5 \times 10^5 $ \msun, about a factor of 2 
lower than derived for a dust temperature of 7 K.

The actual dust mass depends strongly on the choice of
$\beta$. For instance,
if $\beta=1$, we could fit the dust emission longwards
of 60 $\mu$m with only one dust component at a temperature
of about  45 K -- no cold dust being required at all.
The total dust mass would then be much lower, only about $10^4$ \msun.
However, we do not believe $\beta=1$ to be a plausible choice, because  
nearby galaxies provide compelling evidence (see Sect. 4.1)
that $\beta \simeq 2$ for large grains.
Thus, if we had $\beta \simeq 1$  in NGC~1569, it would indicate that
its dust properties are very different from other nearby galaxies,
a possibility no different from the one discussed in Sect. 4.3.

In order to explore the influence of a slightly different
$\beta$, which might still be consistent with 
the observations presented in Sect. 4.1, 
we have fitted the spectrum of \object{NGC~1569} with $\beta =1.8$.
The results are very similar to $\beta = 2$. The hot dust component
is unchanged and the warm and cold component require temperatures of
37 and 7 K, respectively. The total dust mass is slightly reduced 
to $8.7 \times 10^5$ \msun.

In conclusion, if $\beta \simeq 2$ as in all or most other
nearby galaxies, then the dust mass necessary to explain the spectrum
of NGC~1569 within a multi-temperature model
is at least $5 \times 10^5$ \msun and, if we take the
best-fit to the data, even higher, $1.2 \times 10^6$ \msun.          
}

\subsubsection{Where can the cold dust hide?}

Such  a large amount of cold dust 
is a very unlikely state of affairs in a low-metallicity galaxy
dominated by intense radiation fields. For instance, from the UV measurements 
in \cite{israel88} we 
{ estimate a mean 
radiation field at 1000\AA \, within
NGC~1569 (assuming \object{NGC~1569} to be spherical with a radius of 
1 arcmin = 640 pc)  
%
%
of about $1 \times 10^{-16}$ erg sec$^{-1}$ cm$^{-2}$ Hz$^{-1}$.
As the Solar
Neighbourhood is characterised by a value 
%
%
of about $1 \times 10^{-18}$ erg sec$^{-1}$ cm$^{-2}$ Hz$^{-1}$
(Mezger et al. \cite{mezger82}), the 
\object{NGC~1569} radiation field field is two orders of 
magnitude higher }
in good agreement with the 
strong thermal free-free and H$\alpha$ emission from this post-starburst 
galaxy. Such high radiation field intensities suggest that dust temperatures 
should be a factor of $100^{1/5}=2.5$ (for $\beta = 1$) to  $100^{1/6}=2.1$ 
(for $\beta = 2$) higher than for the Solar Neighbourhood. The latter has
to be about 10 K for silicates and 18 K for graphite 
(Mathis et al. \cite{mathis83}) 
which would imply a minimum dust temperature of 20 K in \object{NGC~1569}. The
same conclusion is drawn from a consideration of nearby spiral 
galaxies. Notwithstanding their higher metallicity (more shielding) and
lower radiation field intensities, the coolest component contributing
to their far-infrared/submillimetre spectra has a temperature of typically 
15 K - 20 K (e.g. \object{NGC~891}: Alton et al. \cite{alton}, 
Israel et al. \cite{israel99}; 
\object{M51}: Gu\'elin et al. \cite{guelin}, \object{NGC~3627}: 
Sievers et al. \cite{sievers}).
The coldest dust component in a galaxy such as \object{NGC~1569} with 
very little
shielding against very strong radiation fields should clearly be warmer 
than this. 

{ 
We can estimate the global dust amount from an energy budget
consideration by comparing the 
absorbed to the bolometric radiation. 
Integrating the UV flux densities from 1500 -- 3650 \AA,  the optical 
flux densities from 3650 -- 5500 \AA \ and the infrared-to-millimetre
flux densities from 12 -- 1200 $\mu$m we obtain 
$F_{\rm UV} = 2.9 \times 10^{-9}$, $F_{\rm opt} = 1.7 \times 10^{-9}$ and
$F_{\rm dust} = 3.7 \times 10^{-9}$ erg sec$^{-1}$ cm$^{-2}$.

We have to
make a distinction between the dust emission from H{\sc II} regions and 
the diffuse dust emission. The dust in H{\sc II} regions is mainly heated 
by ionizing photons that are almost completely absorbed by the dust locally, 
practically independent of the dust amount. The diffuse
dust, on the other hand, is mainly heated by nonionizing photons, and the
amount of radiation absorbed is  directly related to the dust opacity. Thus, in
order to calculate the (diffuse) dust opacity, we have to subtract the
dust heating by ionizing photon. We estimate the ionizing UV radiation
from the H$\alpha$ emission.
Waller (\cite{waller}) finds a total H$\alpha$ luminosity of 
$L_{\rm H\alpha} = 4.5 \times 10^{40}$ erg sec$^{-1}$. 
This value is corrected for  the 
Galactic foreground extinction of $E(B-V)=0.56$ (\cite{israel88});
we neglect any 
extinction internal in the H{\sc II} regions. 
The total dust emission originating from ionizing radiation can 
be estimated as $L^{\rm ion}_{\rm dust} = 27.12 \times L_{\rm H\alpha} =
1.2 \times 10^{42}$ erg sec$^{-1}$ (Xu \& Buat \cite{xu95}). 
The corresponding flux is 
$F^{\rm ion}_{\rm dust}=L_{\rm H\alpha}/(4\pi D^2) = 2.0 \times 
10^{-9}$ erg sec$^{-1}$ cm$^{-2}$.

The fraction of radiation absorbed by the diffuse dust is
$(F_{\rm dust} - F^{\rm ion}_{\rm dust})/(F_{\rm UV}+F_{\rm opt}+F_{\rm dust}-
F^{\rm ion}_{\rm dust}) = 0.27 $.  
We estimate the diffuse
dust opacity using a simplified radiation transfer model with a slab geometry,
assuming  that dust and stars are homogeneously mixed.
The dust mass derived in this geometry is higher than if the dust
were in a foreground layer. 
We use an approximate formula (Xu \& de Zotti \cite{xu89}) for the dust
absorption probability,  further simplified by 
applying for the UV and optical radiation the extinction properties
at 2000 and 4300 \AA, respectively (see Lisenfeld, V\"olk \& Xu 
\cite{lisenfeld96}). With this we derive an opacity in the blue of
$\tau_{B} =0.16$. Assuming that the dust has the same extinction properties
as dust in our Galaxy (DBP90), we can estimate the dust mass (in \msun)
as 
\begin{equation}
M_{\rm dust} = 8.8 \times 10^4 \times \tau_B \times 
\left(\frac{A}{\rm kpc^2} \right),
\end{equation}
where A is the surface area of the galaxy. Assuming a circular area
with radius 80$^{\prime\prime}$ we obtain $M_{\rm dust} = 2.3 \times 10^4$
\msun. The uncertainty in this estimate is, even taking into account
unknown parameters like the geometry of dust and stellar distribution, 
certainly less than a factor of 2 so that the dust mass derived is 
more than an order of magnitude lower than the cold dust 
necessary to explain the dust emission spectrum.
If the dust contained more VSG's than in our Galaxy, the derived dust mass
would be lower because the extinction efficiency per mass is higher
for VSG's (DBP90). 

The above considerations do not exclude the presence of very dense 
regions of cold dust 
barely contributing to the extinction or overall emission.
But where could such a dust component be? 
It would have to be in regions of high dust opacity in order to 
be shielded from the interstellar radiation field (ISRF) 
or be situated far away from stars. 
The latter is not the case: \object{NGC~1569} is a small galaxy with an 
intense star-formation  
over the whole area where dust emission is observed. In fact, the
maximum of the dust emission is very close to the large star clusters
A and B (Fig. 1). 
Is the self-shielding of the dust enough to maintain the inner parts cold?
The opacity, $\tau$, along the line of sight can be calculated as:
\begin{equation}
\tau_\nu = \frac{S_\nu}{\Omega B_\nu(T)}
\end{equation}
where $S_\nu$ is the flux density coming from the solid angle $\Omega$.
For a dust temperature of 7 K (11 K) we derive from the peak flux density
of 25 mJy/beam
$\tau_{1200 \mu m} = 1.5 \times 10^{-3}$ ($ 5.8 \times 10^{-4}$).
With $\tau_V/\tau_{1200 \mu m} = 4 \times 10^4$ (Kramer et al. \cite{kramer})
this gives an optical opacity of $\tau_V=58$ (23). 
Maximum shielding could be achieved if this dust were in one big cloud
illuminated from outside only. But also in this case,
a rather thick outside layer of warm  dust exposed to the ISRF
will be present. In order to attenuate the ISRF of \object{NGC~1569}
by a factor of 100, which would make it comparable  to the Galactic ISRF,
a layer of $\tau_V=4.5$ is needed. Thus, we can estimate the  
cold dust mass fraction by
considering a sphere with diameter $\tau_V=58$ (23) where an outer layer 
of  $\tau_V=4.5$ is made of warm dust and the inner
part consists of cold dust. The mass ratio of cold dust to total dust is 
60 \% ($T=7$ K), respectively 20 \% ($T=11$ K). This is not enough to 
explain the dust emission spectrum for which 
97 \% (respectively 94 \% for $T=11$ K) of the dust mass
must be cold.

An alternative way to shield the dust could be molecular gas which
is most likely abundant in the regions of maximum dust emission
(see Sect. 5). However, molecular hydrogen is
dissociated only by photons of energies above 11 eV, corresponding to
radiation shortwards of about 1000 \AA \ so that the overall radiation 
field of \object{NGC~1569} does not get significantly damped.

A final argument against a large amount of cold dust, is that
if it were present,} the resulting dust mass is of order one per cent 
of the gas mass in the same area, $M_{\rm gas} \approx 9.2 
\times 10^{7}$ \msun (\cite{israel88}, \cite{stil01}; 
see also Sect. 4.3.2). This {\it metal-poor} 
galaxy would thus have an usually {\it high dust-to-gas ratio}.

For these reasons, we reject any explanation of the far-infrared/submillimetre
spectrum of \object{NGC~1569} involving the presence of large amounts of cold
dust.

\subsection{An abundance of very small grains?}

\begin{figure}
\resizebox{\hsize}{!}{\rotatebox{270}
{\includegraphics{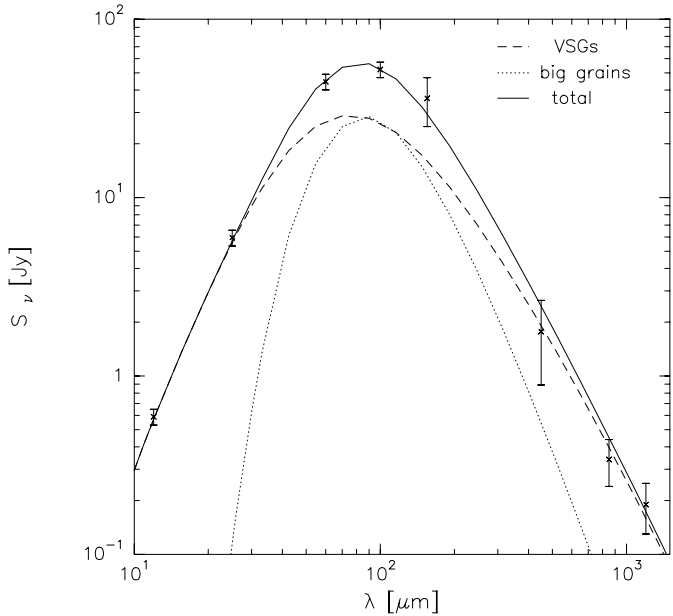}}}
\caption{The mid-infrared to millimetre spectrum of \object{NGC~1569}, fitted
with the dust model of \cite{desert} assuming dust to
be exposed to a radiation field equal to the Solar Neighbourhood field 
scaled up by a factor 60. Emission by PAH's is negligible and the number 
of VSG's is increased by a factor of seven over the local values in 
\cite{desert}.
Crosses mark flux-densities from Table 1.}
\label{spec_isrf}
\end{figure}

As an alternative, we have fitted the dust
 emission spectrum of \object{NGC~1569} with 
the model by \cite{desert}. 
In this model, the dust is described by the three 
components mentioned above (PAH's, VSG's and large grains) exposed to
a radiation field which can be varied in intensity.  
{ The maximum and minimum size, as well as the exponent of the 
size distribution of the various components can be varied. Here, we use
those values for which DBP90 achieved the best-fit for
the solar neighbourhood. 
The big grain size ranges from 15 and 110 nm and  they are distributed
in a power-law with exponent 2.9. The VSGs have sizes between 1.2 and 
15~nm with a power-law exponent of 2.6.  The corresponding values
for PAHs are 0.4 to 1.2~nm  and an exponent of 3. The exponent of the
wavelength dependence of the extinction coefficient is $\beta = 2$
for large grains and $\beta = 1$ for VSGs.}
We have tested 
different radiation fields and allowed for different relative contributions 
of the three components. We obtain a good fit for dust exposed to a 
radiation field similar in spectral shape to the local Solar Neighbourhood 
field, but with an intensity sixty times higher, consistent with the 
{ radiation field of \object{NGC~1569}} estimated in Sect. 4.2.

The resulting fit is shown in  Fig.~\ref{spec_isrf} { and the flux densities
of the various components are listed in Tab. 2}.
Not unexpectedly, the 
abundance of PAH's is found to be negligible because of the steep 
mid-infrared rise in the required spectrum. Madden
(\cite{madden}) reaches a similar conclusion from observations of the MIR/FIR 
spectrum of \object{NGC~1569}. However, the abundance of 
VSG's with respect to the large grains must be increased 
by a factors of about seven 
over the Galactic abundance assumed in the DBP90 model. The
large VSG contribution is required to successfully match both the 
mid-infrared and the submillimetre/millimetre ends of the spectrum. The 
contribution of large grains is well-constrained by the requirement to fit 
the far-infrared data points.

The conclusion that the abundance of VSG's is 
enhanced with respect to the solar
neighbourhood is very robust and
does not depend on the details of the modelling, as we confirmed
by trying to fit the data with other radiation fields given
in DBP90 and changing the size distribution of the VSG's.
Acceptable -- although not equally good -- fits could be
achieved for the radiation field around an O5 star and for an
enhanced UV radiation field. In those case the abundance by
VSG's had to be increased by a factor of 2 -- 3.

Thus, there are two principal differences between our model fit of
\object{NGC~1569} and the model fits presented by DBP90 for the local
Galactic environment. These are: {\it (i) the almost total absence
of PAH emission and (ii) a strong enhancement of the VSG contribution 
relative to that of the large grains by a factor of 2 -- 7}. 
 Both factors point to a different evolution
of dust grains under the different environmental conditions prevalent
in \object{NGC~1569}. 

\subsubsection{Dust grain processing}

The first result is not entirely unique to \object{NGC~1569}. Both 
\object{Magellanic Clouds} 
also suffer from significant PAH depletion but not as strongly 
as \object{NGC~1569} (Schwering \cite{schwering}; Sauvage et al. 
\cite{sauvage}). However, it is
relevant to note that, although the metallicity of \object{NGC~1569} 
($12+\log(O/H) = 8.19$, Kobulnicky \& Skillman \cite{kobulnicky}) is in
between 
those of the\object{LMC} and the \object{SMC}, 
the intensity of its radiation field is 
5 -- 10 times higher. PAH depletion, in fact, appears to be a general
characteristic of dwarf galaxies. In the compilation by Melisse $\&$ Israel
(\cite{melisse}), 
\object{NGC~1569} clearly has the highest $f_{25}/f_{12}$ ratio (about 10),
but it is followed by two more galaxies (\object{NGC~3738} and 
{\object{IC~4662}) with ratios of 7
and 8 respectively. The mean $f_{25}/f_{12}$ ratio of the remaining 
ten Im galaxies is $4.5\pm0.5$, whereas 15 larger galaxies classified
Sm have a mean ratio $2.9\pm0.4$. Large late-type spiral galaxies with
less intense mean radiation fields and higher metallicities mostly have
ratios 1.5 -- 2. With the $f_{25}/f_{12}$ ratio as indicator, irregular 
dwarf galaxies clearly have dust contents significantly depleted in PAH's
as compared to spiral galaxies. Most likely, PAH depletion is a function 
of the incident UV radiation, i.e. of the global UV radiation intensity 
and the amount of shielding available (Puget \& L\'eger \cite{puget}).
In this respect, \object{NGC~1569} provides
one of the more extreme environments by being metal-poor {\it and}
presently in a (post) starburst phase.

The magnitude of the increase in the VSG/large grain ratio with respect
to that of the Solar Neighbourhood, suggests the amount of 
VSGs increase at the expense 
of large grains by extensive processing of the latter. 
Grain-grain collisions in slow
shocks can efficiently destroy large grains by
fragmenting and shattering and convert them into small grains
producing an excess of small particles as we believe to see in
\object{NGC~1569} (Borkowski \& Dwek \cite{borokowski}, Jones  et al. 
\cite{jones}).
\object{NGC~1569} is a post-starburst galaxy. A large
fraction of its interstellar medium is in chaotic motion with a rather
high {\it mean} velocity dispersion of about 20 km s$^{-1}$, and
over a dozen radio supernova remnants can be identified in this small
galaxy. Thus,  shocks  
are necessarily present in its interstellar medium. 
In addition, the low metallicity of \object{NGC~1569} may also hamper formation
of large grains by inefficient accretion to begin with.

\subsubsection{Dust mass and gas-to-dust ratio}

The total dust mass derived from the adapted \cite{desert} model is much less
than that found for the model based on cold large grains discussed
in the preceding section. 
{ The mass of the large grains is derived from eq.~(\ref{mdust})
with $S_{100 \mu m}$ and $T$ being the flux density at 100 $\mu$m of 
large grains and their temperature.}
The mass of the VSG component is found from
the relative mass contributions of VSG's and large grains given by 
\cite{desert} (their Table 2), scaled up by the factor 7 that we have found for
\object{NGC~1569}. Notwithstanding the substantial number 
increase of small grains 
with respect to large grains, the total dust mass $M_{\rm d,tot} = 3.2 
\times 10^{4}$ \msun \ is still dominated by the large grains, with 
$M_{\rm d,LG} = 2.1 \times 10^{4}$ \msun \ (66$\%$). 
The large-grain temperature required by the model fit ($T_{LG}$ = 32 K) is 
close to the temperature ($T_{\rm IRAS} \approx$ 35 K) that, ignoring 
everything else,  we would obtain directly from the IRAS $f_{60}/f_{100}$ 
flux-density ratio. Partly because of this, the dust mass derived here is 
somewhat fortuitously not very different from the mass $2.8 \times 10^{4}$ 
\msun we would naively obtain from the IRAS 60$\mu$m and 100$\mu$m data 
only. This situation is very different indeed from the one pertinent to 
spiral galaxies where IRAS-derived dust masses commonly underestimate the 
total dust mass by an order of magnitude (Devereux \& Young \cite{devereux}).
{ The uncertainty in this mass derivation can be estimated by
considering  the masses derived for the full range of fits 
consistent with the errors. In no case a lower dust mass was derived.
The highest dust mass derived (in a cases where a lower VSG abundance
was needed) was $6.3 \times 10^4$ \msun, i.e.  not even  a factor of
2 higher. The range in dust mass derived from the DBP90 model
is thus $3.2 - 6.3 \times 10^4$ \msun.}

We have summed HI maps of \object{NGC~1569} (\cite{stil01}) over 
a circular area with radius of 80\arcsec \, comprising all the
flux in the 1200$\mu$m map. We find an HI mass of 
$M_{\rm HI} = 5.1 \times 10^{7}$ \msun.
For mass contributions $M_{\rm H2} = 0.35 M_{\rm HI}$ 
(Israel \cite{israel97}) and $M_{\rm He} = 0.25 M_{\rm gas}$,
and a dust mass $M_{\rm d} = 3.2 -  6.3 \times 10^4$ \msun ,
we find an overall gas-to-dust ratio $M_{\rm gas}/M_{\rm d} \approx 
1500 - 2900$ which is significantly higher than the Solar Neighbourhood ratio. 
The corresponding ratio $M_{\rm HI}/M_{\rm d} \approx 1600$ is in good 
agreement with the prediction by Lisenfeld $\&$ Ferrara (\cite{lisenfeld}) 
that 
this ratio should exceed the Solar Neighbourhood value typically by an 
order of magnitude for metallicities 12+log(O/H) = 8.70 and 8.19 in the 
Milky Way and \object{NGC~1569} respectively. 

\section{Distribution of dust, atomic and molecular gas}

The (sub)millimetre maps not only closely resemble one another, but also 
maps of centimetre-wavelength radio continuum emission 
(Israel \& de Bruyn \cite{israel-bruyn}, Wilding \cite{wilding}) 
and H$\alpha$ (Waller \cite{waller}) images. In 
all these maps emission peaks in the western part of the optical image 
of the galaxy, just west of `super' star cluster A (see Fig. 1). 
The 850$\mu$m/1200$\mu$m 
continuum peak is close to the discrete CO clouds mapped by Greve et al. 
(\cite{greve}) and Taylor et al. (\cite{taylor99}), 
but offset from these by about 10$''$ 
to the east.  In all maps major emission curves eastwards, north of star 
clusters A and B, following a string of optically prominent HII regions 
(cf. Waller \cite{waller}). 
The overall extent of the (sub)millimetre continuum 
emission is very similar to the optical extent, and that at radio 
continuum wavelengths.

We might suspect that contamination of the (sub)millimetre continuum maps 
by free-free emission from ionized gas causes their resemblance to 
radio continuum and H$\alpha$ maps. Indeed, in Sect. 3. we have shown that 
our maps at 1200$\mu$m and at 850$\mu$m contain non-negligible contributions 
of free-free emission. However, the fraction of global emission involved
(25$\%$ and 13$\%$ respectively) is insufficient to explain the resemblance
especially at 850$\mu$m. Nevertheless, the question thus arises whether 
the free-free emission emission is proportional to the dust emission 
over the entire maps. If this is not the case, specific structural details 
in our maps could indeed be due to local concentrations of free-free 
emission. In order to address these concerns, we have compared the 
H$\alpha$ map by Waller (\cite{waller}; his Fig. 3a) to our maps. 
The H$\alpha$ 
contour values multiplied by factors of $5.1 \times 10^{-14}$ and $4.9 
\times 10^{-14}$ should correspond to the free-free emission contribution
across \object{NGC~1569} in our 1200$\mu$m and 850$\mu$m maps. With these conversion 
factors, the H$\alpha$ map shows that the free-free emission from 
ionized gas is on the
whole reasonably well scaled with the thermal emission from dust. A notable 
exception is the main emission peak at $\alpha$(2000) = $4^{h}30^{m}47^{s}$, 
$\delta$(2000) = $+64^{\circ}50^{'}58{''}$, coincident with the 
most prominent HII region complex of \object{NGC~1569}
(HII 2 in Fig. 1). The ionized gas should 
contribute about 12 mJy/beam to the peak flux densities in our maps. 
This corresponds to disproportionate fractions of about 50$\%$ and 20$\%$ 
at 1200$\mu$m and 850$\mu$m respectively, about twice the global average.
Below emission levels of 15 - 20 mJy/beam (1200$\mu$m) and 40 - 50 mJy/beam
(850$\mu$m) ionized gas and dust emission are roughly proportional,
equal to the global average.

\begin{figure}
\resizebox{\hsize}{!}{\includegraphics{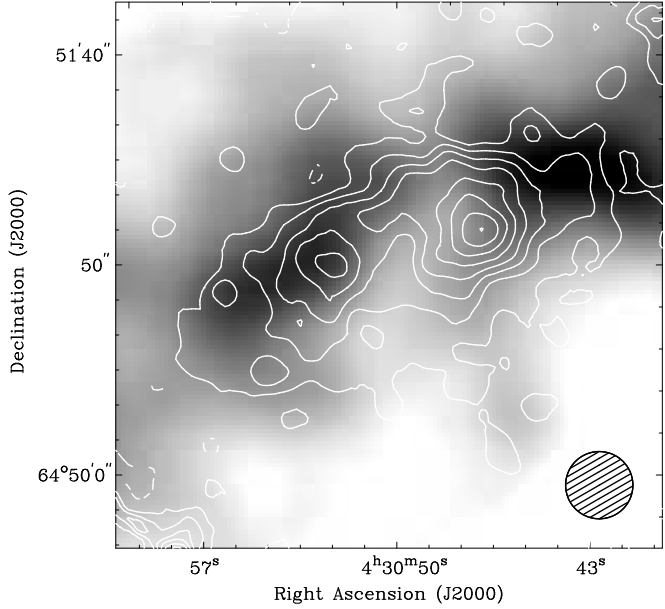}}
\caption{Intensity at 850$\mu$m (contours) over HI column density
(greyscales).
Contour levels are $-$8, 8, 16, 24, ... mJy/beam. 
Grayscales increase linearly
from $1 \times 10^{21}$ to $7 \times 10^{21}\ \rm cm^{-2}$. The hatched
circle 
indicates the size of the $16"$ beam (FWHM) applicable to both maps.
}
\label{hi}
\end{figure}

\begin{figure}
\resizebox{\hsize}{!}{\includegraphics{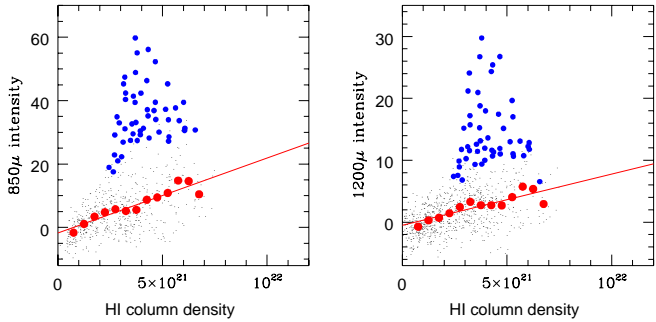}}
\caption{Correlation between HI column density and intensity at 850$\mu$m 
(left) and 1200$\mu$m (right). Small and medium-sized black dots 
correspond to 5$''$.  The medium-sized dots mark positions where the 
21-cm radio continuum intensity is higher than 20 mJy/beam. The large
dots are binned averages of the remaining points (small dots) and the 
straight line is a fit to these average points.
}
\label{column}
\end{figure}

We have convolved to the 16$''$ JCMT 850$\mu$m resolution not only the
IRAM 1200$\mu$m map, but also the WSRT HI column density map (\cite{stil01}) 
(Fig.~\ref{hi}). 
The (sub)millimetre emission is mostly just south of the HI ridge
in \object{NGC~1569}. The main peak, marking the location of molecular clouds
and bright HII regions, is located on a saddle point in the HI distribution,
i.e. close to a local minimum in the HI ridge, between the two
elongated major HI maxima. We may use the maps for a direct comparison of 
the (sub)millimetre emission
intensity with the column density of neutral hydrogen (HI). 
Fig.~\ref{column} displays the correlation between the latter and 
the 850$\mu$m and 1200$\mu$m intensities. The overwhelming majority of 
points (small dots in Fig.~\ref{column}) follows a linear relationship 
between broadband intensity and HI column density. At these intensity 
levels free-free emission is a more or less constant fraction of dust 
emission, and at these wavelengths most of the continuum emission is
contributed by very small grains. Thus, the linear relation reflects a 
fairly constant VSG-to-gas ratio in \object{NGC~1569}. The relatively small
fraction of points that deviate strongly from the general correlation 
is contributed entirely by the regions of most intense emission in the 
SCUBA/IRAM maps, notably the peak just mentioned. 

In order to determine the general correlation, we first applied a cutoff
for all points with more than 20 mJy/beam in the 21-cm {\it radio 
continuum} map; this effectively excludes all strongly deviating points.
We then averaged the dust emission in $5\times 10^{20}\ \rm cm^{-2}$ intervals 
of HI column density. The resulting points are marked by large dots
in Fig.~\ref{column}.  A fit to the points thus found, corrected for
the free-free contribution, yields the relations $I_{1200} = 6.2 \times 
10^{-22} N_{\rm HI} -0.51$ and $I_{850} = 2.0 \times 10^{-21} N_{\rm HI} 
-1.71$, ($N_{\rm HI}$ in $\rm cm^{-2}$, $I$ in mJy per 16 arcsec beam). 
Assuming a constant gas-to-dust mass ratio, and neglecting local temperature 
variations, the deviating points then seem to imply the presence of a large 
amount of undetected hydrogen, which should be in molecular form. This 
consideration allows us to estimate the molecular mass independently 
from uncertain CO-to-$\rm H_2$ conversion factors (Israel \cite{israel97}).

The largest difference is found for the peak in the maps, with a {\it mean} 
intensity 46 mJy per 16 arcsec beam at 850$\mu$m and 22 mJy per
beam at 1200$\mu$m.
Corrected for a mean free-free-contribution of
10 mJy/beam (somewhat less than the free-free peak), these intensities 
imply total hydrogen column densities $N_{\rm H} = 2 \times 10^{22} 
\rm cm^{-2}$. The difference $N_{\rm H} - N_{\rm HI}$ is the column density 
of hydrogen atoms bound in molecules. Thus, Fig.~\ref{column} suggests
a mean molecular hydrogen column density $N_{\rm H_2} = 8 \cdot 10^{21} \rm 
cm^{-2}$. 

Division by the mean integrated CO intensity over this area,
$I_{\rm CO} \approx 1.5$ K km s$^{-1}$ (Greve et al. \cite{greve}) yields an
estimate for the actual CO-to-$\rm H_2$ conversion factor $X_{1569}
\approx 5.3 \times 10^{21}$ cm$^{-2}$ (K km s$^{-1}$)$^{-1}$. This
result, about 25 -- 30 times the commonly adopted Galactic value, is very
close to that obtained by Greve et al. (\cite{greve}), but rejected by them;
it is a factor of three below the high but uncertain value estimated
by Israel (\cite{israel97}) 
and about a factor of three higher than the estimate
by Taylor et al. (\cite{taylor99}) 
which was derived from virial considerations and 
therefore should be considered a lower limit under the conditions
prevalent in \object{NGC~1569} (Israel \cite{israel97}; \cite{israel00}).

{ It is interesting to note that the deviating points in Fig.~\ref{column},
implying the need of molecular gas, only occur for 
$N_{\rm HI} \geq 3 \times 10^{-21}$ cm$^{-2}$. This could be due to a 
threshold in $N_{\rm HI}$ necessary for the formation of molecular gas.
Such a threshold exists in our Galaxy around 
$N_{\rm HI} \approx 5 \times 10^{-20}$ cm$^{-2}$ 
(Federman et al. \cite{federman}) and reflects the   
balance between molecular gas formation and destruction by 
photodissociation. A minimum gas column density
is necessary in order to protect the molecular gas from the
radiation field through self-shielding 
as well as shielding by dust grains. It can be expected that 
in \object{NGC~1569}  the threshold $N_{\rm HI}$
is higher, in agreement with our finding,
because the low dust content decreases both the shielding
and the molecular gas formation rate and because the high radiation
field increase the molecular gas destruction rate.}

As can be seen in Fig.~\ref{hi}, adding these molecular 
hydrogen column densities to those of HI results in a total filling-in
of the HI ridge minimum. It appears that the total hydrogen distribution in
\object{NGC~1569} remain ridge-like, but with a smooth increase to a pronounced
maximum coincident with the present continuum peak. 

\section{Conclusions}

\begin{enumerate}

\item We present new maps of the dwarf galaxy \object{NGC~1569} at 450, 850 and
1200 $\mu$m taken with SCUBA at the JCMT and the MPIfR bolometer array 
at the IRAM 30m telescope. Integrated flux-densities at these wavelengths
may be compared to those at at 12, 25, 60 and 100 $\mu$m obtained
earlier with IRAS.

\item The steep rise in intensity from 12 to 25 $\mu$m excludes
a significant contribution from PAH's, as the broadband spectrum
of these decreases in wavelength.

\item The (sub)millimetre flux densities are high compared to the flux 
densities at shorter wavelengths and the (sub)millimetre spectrum 
has a relatively shallow slope ($\simeq \lambda^{-2.5}$). Such a spectral 
shape can be explained by the presence of a significant amount of cold 
dust. Fits to the observed spectrum by three-temperature dust models,
however, require most dust to be at temperature of only about 7 K. 

\item We do not favour this explanation. The intense radiation field
and low metallicity of \object{NGC~1569}, implying poor shielding, render
it very unlikely that large amounts of dust at such low temperatures
may exist in \object{NGC~1569}. { We show that
the high dust opacities necessary
to shield a large fraction of the dust from this radiation field
are not present.}
Even if it could, the resulting gas-to-dust mass 
ratio would have to be about 100, again an unusually low value for a 
low metallicity galaxy such as \object{NGC~1569} ($12+\log(O/H)=8.19$).

\item Alternatively, the (sub)millimetre flux densities may be dominated
by emission of VSG's at various non-equilibrium temperatures. The combined
emission is characterized by a wavelength dependence mimicking an
extinction coefficient $k_\lambda \propto \lambda^{-1}$. Using the
dust model of \cite{desert}, we achieve good fits for dust exposed to
radiation fields similar in spectral shape to the Solar Neighbourhood 
field but with sixty times higher intensity, as is appropriate for
\object{NGC~1569} (\cite{israel88}). 

\item The fits require VSG to large grain 
abundances to be enhanced by a factor of 7 
as compared to those in Solar Neighbourhood. The precise value of the 
enhancement factor is slightly model dependent; use of different radiation 
field yielding less good, but still acceptable fit results, yields
slightly different factors. A robust conclusion is that the VSG
enhancement factor in \object{NGC~1569} is 2 -- 7.  

\item Although in these models much of the emission originates from
very small grains, virtually all of the mass still resides in 
the large grains. The gas-to-dust mass ratio is 1500 -- 2900, about
an order of magnitude higher than in the solar neighbourhood.

\item Both the dust and molecular gas distributions peak at a local 
minimum in the HI ridge. If gas-to-dust mass ratio are constant over
\object{NGC~1569}, the lack of HI at the very peak of the dust emission 
indicates 
the presence of a significant column of molecular gas. From the inferred
molecular gas column density and the observed CO emission, we estimate 
the hydrogen column density to integrated CO intensity conversion factor
$X \approx 5 \times 10^{21} $ H$_{2}$ mol cm$^{-2}$ (K km s$^{-1})^{-1}$,
or about 25 -- 30 times the local Galactic value.
\end{enumerate}

\begin{acknowledgements}

We would like to thank F.--X. D\'esert for providing us with his
program of the dust model, Martijn Kamerbeek for help in the data 
reduction and the JCMT staff, in particular Fred Baas, for carrying 
out the observations in service mode. { We would like to thank the
anonymous referee for many useful suggestions.}
This work made use of the NASA Extragalactic Database.

\end{acknowledgements}

\end{document}